\documentclass[prx, aps, longbibliography, twocolumn, reprint, showpacs, footnoteinbib, superscriptaddress]{revtex4-1}
\usepackage{graphicx}
\usepackage{amssymb}   
\usepackage{amsfonts}
\usepackage{amsmath}
\usepackage{dsfont}
\usepackage{dcolumn}   
\usepackage{bm}        
\usepackage[mathscr]{eucal}
\usepackage[dvipsnames]{xcolor}
\usepackage[colorlinks, linkcolor=OrangeRed,citecolor=RoyalBlue,urlcolor=NavyBlue]{hyperref}
\usepackage[all]{hypcap} 
\usepackage{mathtools}
\usepackage{wasysym}
\usepackage{tikz}
\definecolor{redpaper}{rgb}{0.69, 0.18, 0.24} 
\definecolor{bluepaper}{rgb}{0.08, 0.34, 0.62} 
\def\bea{\begin{eqnarray}}
\def\eea{\end{eqnarray}}
\newcommand{\beq}{\begin{equation}}
\newcommand{\eeq}{\end{equation}}

\begin{document}

\title{Real-time dynamics of string breaking in quantum spin chains}
\author{Roberto Verdel}
\affiliation{Max-Planck-Institut f\"ur Physik komplexer Systeme, N\"othnitzer Stra{\ss}e 38,  01187-Dresden, Germany}
\author{Fangli Liu}
\author{Seth Whitsitt}
\author{Alexey V. Gorshkov}
\affiliation{Joint Quantum Institute, NIST/University of Maryland, College Park, Maryland 20742, USA}
\affiliation{Joint Center for Quantum Information and Computer Science, NIST/University of Maryland, College Park, Maryland 20742, USA}
\author{Markus Heyl}
\affiliation{Max-Planck-Institut f\"ur Physik komplexer Systeme, N\"othnitzer Stra{\ss}e 38,  01187-Dresden, Germany}

\begin{abstract}
String breaking is a central dynamical process in theories featuring confinement, where a string connecting two charges decays at the expense of the creation of new particle-antiparticle pairs. Here, we show that this process can also be observed in quantum Ising chains where domain walls get confined either by a symmetry-breaking field or by long-range interactions. We find that string breaking occurs, in general, as a two-stage process: First, the initial charges remain essentially static and stable. The connecting string, however, can become a dynamical object. We develop an effective description of this motion, which we find is strongly constrained. In the second stage, which can be severely delayed due to these dynamical constraints, the string finally breaks. We observe that the associated time scale can depend crucially on the initial separation between domain walls and can grow by orders of magnitude by changing the distance by just a few lattice sites.  We discuss how our results generalize to one-dimensional confining gauge theories and how they can be made accessible in quantum simulator experiments such as Rydberg atoms or trapped ions.  
\end{abstract}

\maketitle
\section{Introduction} \label{Introduction}

Confining theories such as quantum chromodynamics have the defining property that two static charges, e.g., a heavy quark-antiquark pair, are connected by a flux tube or \textit{string}, whose energy increases linearly with the separation \cite{Alkofer_2007}. Beyond some critical distance, however, the string can break as the creation of new, light particle-antiparticle pairs becomes more favourable \cite{GLIOZZIA199976, GLIOZZI, Philipsen1998, KNECHTLI1998345}. This  mechanism is known as \textit{string breaking} and has been investigated extensively from a static point of view~\cite{Philipsen1998, PHILIPSEN1999146, KNECHTLI1998345, KNECHTLI2000309, L_scher_2002, GLIOZZI200591, BaliPRD, Pepe2009} while recently also its dynamics has gained increased attention~\cite{ Gelfand, Banerjee2012, StringBreakingMPS,  PhysRevX.6.011023, KASPER2016742, QEDMPS, kuno2017, Sala2018PRD, Spitz&Berges, park2019, PhysRevX.10.021041, Magnifico2020realtimedynamics, PhysRevResearch.2.013288, PhysRevLett.124.180602}. Importantly, many aspects of confinement cannot only be realized in gauge theories, but also in conventional quantum spin chains \cite{mccoy, Greiter2002, Rutkevich2008, Lake, Coldea177, Bose2010, PhysRevB.83.020407, Cai2012, Morris2014, Grenier2015, Wang2016, Bera, calabreseNat, Diamantini2018, Long-range, Mazza, SilvaLongRange, Sulejmanpasic2017, Gannon2019, PhysRevB.101.115138}. Yet, it has remained an open question whether quantum spin models can inherit also the fundamental dynamical process of string breaking.

\begin{figure}[tb!]
    %\centering
	\includegraphics[width=1.\columnwidth]{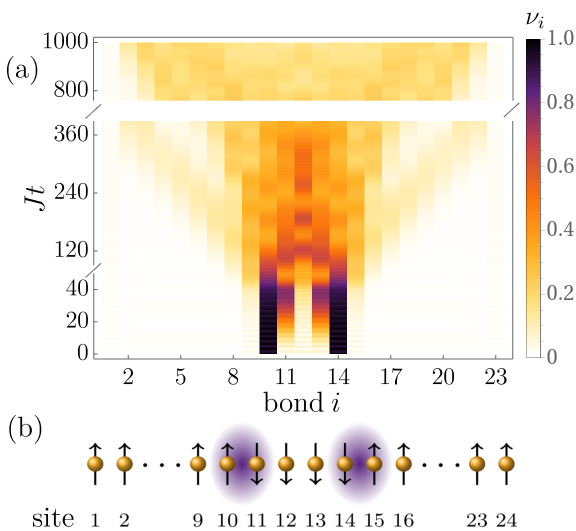}
	\caption{Real-time dynamics of string breaking in the short-range Ising chain with $L=24, h_x/J=0.2, h_z/J=1$ and an initial distance $\ell=4$ between two domain walls.  (a) Dynamics of the domain wall density $\nu_i(t)$ displaying two stages of string breaking. First, the initial domain walls remain static for a long time  with dynamics occurring in the connecting string. Second, the string breaks on longer time scales by forming bound pairs of domain walls. The initial string state is schematically depicted in (b).}
	\label{fig:1}
\end{figure}

In this work we address this question and show that string breaking can occur in paradigmatic quantum Ising chains. Here, the elementary excitations are domain walls which can exhibit confining potentials induced either by symmetry-breaking fields~\cite{mccoy, Rutkevich2008, calabreseNat} or long-range interactions~\cite{Long-range, SilvaLongRange}. As a particular consequence, the phenomenology of string breaking not only obtains a significantly broadened scope towards the realm of quantum many-body theory but also brings it within reach of experiments in quantum simulators such as systems of Rydberg atoms or trapped ions. We find that string breaking takes place as a two-stage process (see Fig.~\ref{fig:1}). In the first stage, the two initial charges remain essentially static and stable on a time scale which can depend crucially on the initial domain wall separation. In this regime, we observe, however, that the connecting string can become a dynamical object. We develop an effective description for this string motion, which turns out to exhibit strong kinetic constraints.  The resulting reduced models allow us to obtain analytical access for instance on time scales of string breaking or on bounding the maximum number of particle-antiparticle pairs created during string motion. We further observe that the string motion also leads to a heterogeneous spatiotemporal profile of quantum correlations. While the regions outside of the string essentially remain uncorrelated, the string itself can develop strong quantum correlations. In the second stage, the string eventually breaks at a time scale, which can grow by orders of magnitude upon increasing the separation of the initial domain walls. While we present our findings for two particular quantum Ising chains, we argue that our observations also generalize to other systems such as one-dimensional confining gauge theories. We further discuss how our results on string breaking can be realized in systems of Rydberg atoms and trapped ions.

The structure of this paper is as follows. In Sec.~\ref{Models}, we start by introducing the model Hamiltonians that we consider, and by defining the quench protocols and measured observables.  We present a summary of our main results in Sec.~\ref{summary}. We then further elaborate on these results in Secs.~\ref{first} and \ref{secondStage}. In particular,  in Sec.~\ref{first}, we analyze the first stage of string breaking, presenting some effective descriptions for the string motion during this stage (Secs.~\ref{effective} and \ref{resonant_subspace}), and a bound on the maximum charge density that can be created (Sec.~\ref{part_prod_bound}). The second stage is discussed in Sec.~\ref{secondStage}. Some concluding remarks, including possible experimental implementations of the phenomenology studied in this work, are given in the last section.

\section{Models and quench dynamics} \label{Models}

\subsection{Quantum Ising chains} \label{Hamiltonians}

We study the real-time dynamics of string breaking in two  quantum spin models with distinct features. In the first place, we consider a quantum Ising chain with nearest-neighbor interactions in both transverse and longitudinal magnetic fields, with strength $h_x$ and $h_z$, respectively,
\begin{equation}
	H_{\text{short}}=-J\sum_{i=1}^{L-1} \sigma_i^z\sigma_{i+1}^z-h_x\sum_{i=1}^{L} \sigma_i^x-h_z\sum_{i=1}^{L} \sigma_i^z.
	\label{Ising_short}
\end{equation}
The second model is a quantum spin chain with long-range interactions,
\begin{equation}
	H_{\text{long}}=-J\sum_{i < j}^{L} \frac{1}{r_{ij}^{\alpha}} \sigma_i^z\sigma_{j}^z-h_x\sum_{i=1}^{L} \sigma_i^x,
	\label{Ising_long}
\end{equation}
where $\sigma^{\mu}_i$ ($\mu = x, y, z$) denotes the Pauli matrices acting on site $i$, $r_{ij}$ is the distance between sites $i$ and $j$, $\alpha > 1$ determines the power-law decay of the long-range interactions (the case $\alpha \in [0,1]$ is avoided so as to ensure a well-defined thermodynamic limit \cite{Long-range, Cannas1996}), $L$ is the size of the system, and the ferromagnetic coupling $J > 0$ sets the overall energy scale. For the short-range model in Eq. (\ref{Ising_short}), we use open boundary conditions, since this choice resembles  the  conditions in relevant experimental platforms, which are discussed subsequently. We have, however, checked that our results do not rely on this particular choice.  On the other hand, we consider periodic boundary conditions for the long-range model in Eq. (\ref{Ising_long}), where $r_{ij} = \min \left(|i-j|, L-|i-j|\right)$, as this model is more sensitive to finite-size effects.
Also, throughout this work, we use units such that the reduced Planck's constant $\hbar$ and the lattice spacing $a$ are both set to 1. 

In their respective ground states, both models feature a ferromagnetic phase for sufficiently weak transverse fields $h_x$, with domain walls as the elementary excitations. In the short-range model, this is the case in the limit of vanishing longitudinal field $h_z$, whereas for the system in Eq.~(\ref{Ising_long}) when the power-law decay $\alpha>2$ of the interactions is sufficiently rapid. Upon adding $h_z$~\cite{mccoy, Rutkevich2008, calabreseNat} or upon decreasing $\alpha$ into the range $\alpha<2$~\cite{Long-range}, the domain walls develop a confining potential with the interaction energy between two domain walls increasing as a function of their distance similar in phenomenology to confinement in gauge theories. 

In spite of their similarities, we also note important conceptual differences between the two models. The short-range model has a two-fold degenerate ground state at $h_z = 0$, which is split by the addition of the longitudinal field. Then, the domain wall excitations are points along the chain where the spin tunnels between the two ground states. Since one of the ground states is now higher in energy, two domain walls separated by a string of length $\ell \gg 1$ have an energy cost $E$ proportional to $\ell$. %$E \propto \ell$. 
In contrast, the ground state of $H_{\text{long}}$ is always exactly two-fold degenerate in its ferromagnetic phase, but confinement between domain walls is driven by \emph{frustration} between segments of the chain with opposite magnetization and long-range ferromagnetic couplings. For the Hamiltonian in Eq.~\eqref{Ising_long}, the energy cost of separating two domain walls a distance $\ell \gg 1$ scales as $E \propto \ell^{2 - \alpha}$ ($\log \ell$ for $\alpha = 2$)~\cite{Long-range}. Therefore, the long-range model can interpolate between logarithmic and linear confinement, which are both realized in lattice gauge theories \cite{KogutRMP};  yet we will see that string breaking proceeds in most of the aspects similarly in both models, suggesting that the developed picture is general for theories featuring confinement.

Finally, let us also notice that previous works have explored  connections between Ising models and confining field theories. In particular, duality transformations have been established~\cite{Wegner1971, KogutRMP, PhysRevD.11.2098, PhysRevD.19.3715} between short-range Ising models and some lattice  gauge theories. Similar dualities are expected to hold for the long-range case. Moreover, in Ref.~\cite{Shankar05} it is discussed how the short-range model in Eq.~\eqref{Ising_short}, could be mapped into a gauge theory and  be used to describe the low-energy  physics of the one-dimensional massive Schwinger model. Additionally, there is a recent concrete theoretical proposal~\cite{ZhangPRL2018} to realize certain lattice gauge theories using the system~\eqref{Ising_short}, in the context of quantum simulators. Thereupon,  quantum Ising chains do constitute reasonable lattice theories to study the dynamics of confinement and string breaking.

%Let us also remark that although quantum Ising chains do not constitute lattice gauge theories, they are certainly not completely unrelated. In particular, the  short-range model has been linked to gauge theories featuring confinement\cite{KogutRMP, PhysRevD.11.2098, Shankar05}. 

\subsection{Quantum quench and measured observables} \label{quench}

\begin{figure*}[bt!]
 	\centering
 		\includegraphics[width=\textwidth]{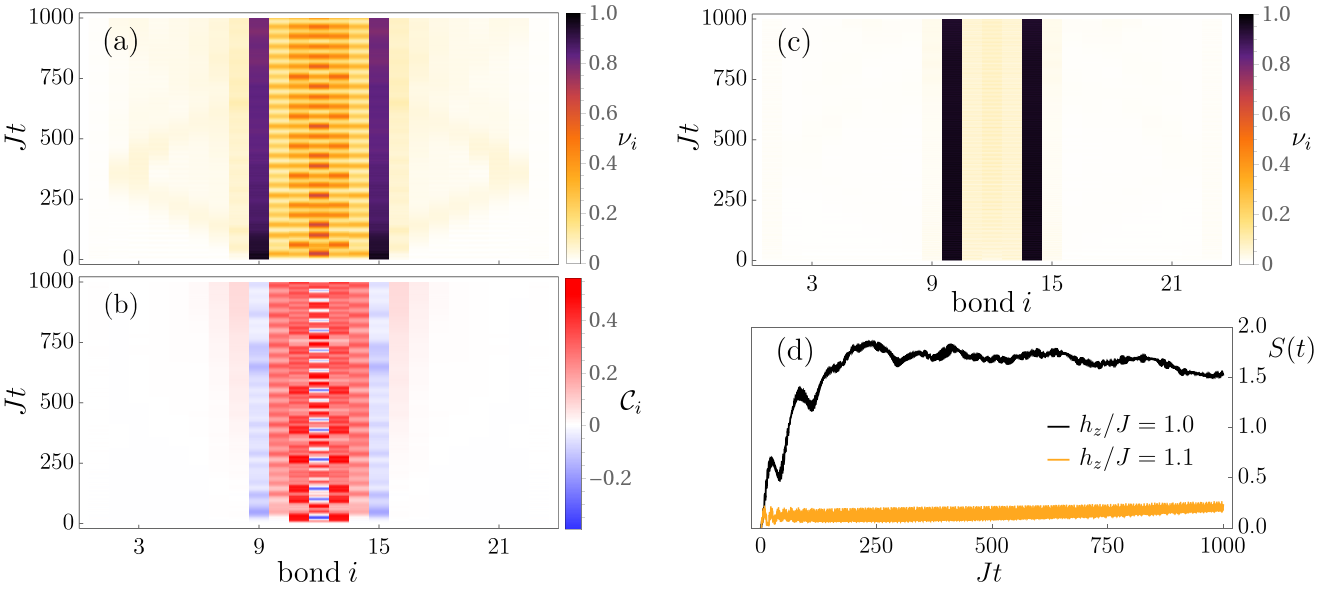}
 	\caption{(Left) String motion in the short-range model with $L=24$, $h_x/J=0.2, h_z/J=1$, and  an initial separation of the domain walls $\ell=6$. Dynamics of (a) the domain wall density $\nu_i(t)$, and (b) the nearest-neighbor connected correlator $\mathcal{C}_i(t)$. (Right) Suppression of string motion and string breaking in the short-range model with $L=24$, $h_x/J=0.2, \ell=4$. Dynamics of (c) the domain wall density $\nu_i(t)$ with  $h_z/J=1.1$, and  (d) the half-chain entanglement entropy $S(t)$ with $h_z/J=1$ (resonant) and $h_z/J=1.1$ (off-resonant). Similar off-resonant behavior is also observed with values of $h_z/J$ smaller than 1. The resonant curve in (d) corresponds to the quench displayed in Fig.~\ref{fig:1}.}
 	\label{fig:2}
\end{figure*}

We study the dynamics of string breaking by initializing the spin chains in a product state with a specific magnetization profile, as shown in Fig.~\ref{fig:1}(b). All spins are pointing $\uparrow$ except within a central region of variable length $\ell$ where the spins are taken to be $\downarrow$. This generates a state with exactly two domain walls connected by a string. This setup not only represents a direct realization of the desired particle-antiparticle pair but is also motivated by the classes of initial conditions that can be prepared experimentally in quantum simulators such as Rydberg atoms or trapped ions; see, for instance~\cite{Marcuzzi2017, 2017Jurcevic, 2017Monroe53}. Let us note that this type of string-like states, as well as excitations with a larger number of strings, have been found to also play an important role in quantum spin dynamics beyond confinement, in other models of quantum magnetism such as the one-dimensional spin-1/2 Heisenberg model~\cite{Wang2018,PhysRevB.100.184406}.

Next, the system is evolved with one of the Hamiltonians in Eqs.~\eqref{Ising_short} or \eqref{Ising_long}. In either case, the transverse field is chosen sufficiently weak so that the elementary domain wall excitations are still almost point-like particles. This setup represents a quantum quench from an excited eigenstate beyond the ground state manifold in the limit of $h_x=0$ to the respective quantum Ising models.  In general, we solve this dynamical problem by means of exact diagonalization (ED) techniques supported by effective analytical descriptions that will be presented in more detail below. Note that, in the regime of stronger transverse fields, domain walls cannot simply be treated as point-like particles; instead, they become extended objects. We provide a brief discussion on the dynamics of kinks with a finite width in Appendix \ref{AppendixE}.

We characterize the resulting dynamics through different observables. On the one hand, we study the dynamics and creation of the elementary excitations by computing the local density of kinks
\begin{equation}
\nu_i(t)=\frac{1}{2}\langle 1 - \sigma_i^z(t)\sigma_{i+1}^z(t)\rangle,
\label{density}
\end{equation}
measuring the presence or absence of a domain wall at the given bond $(i,i+1)$.

Further, we aim to explore the spatiotemporal structure of quantum correlations during string breaking dynamics. For that purpose we study the nearest-neighbor connected correlation function:
\begin{equation}
\mathcal{C}_i(t)=\langle \sigma_i^z(t)\sigma_{i+1}^z(t)\rangle -\langle \sigma_i^z(t)\rangle\langle \sigma_{i+1}^z(t)\rangle.
\label{corr}
\end{equation}
  
Lastly, we also quantify quantum correlations by looking at the half-chain entanglement entropy. To compute this quantity, we partition the system across its center, such that the two resulting subsystems $A$ and $B$, are the left and right halves of the chain, respectively. Then,  the half-chain entanglement entropy is given by the von Neumann entropy of one of the two parts \cite{Tagliacozzo2008}, say $A$, that is,   

\begin{equation}
S(t)\equiv S(\rho_A(t))= -\mathrm{Tr}_A (\rho_A(t) \ln\rho_A(t)),
\label{entanglement}
\end{equation}
\noindent where $\rho_A(t)=\mathrm{Tr}_B (|\Psi(t)\rangle\langle\Psi(t)|)$ is the reduced density matrix of the left half of the chain, and $|\Psi(t)\rangle$ describes the (pure) quantum state of the entire system at time $t$. This entanglement entropy measures the amount of quantum correlations established between the two halves of the chain.

\section{Summary of main results} \label{summary}
 
\begin{figure*}[bt!]
 	\centering
 		\includegraphics[width=\textwidth]{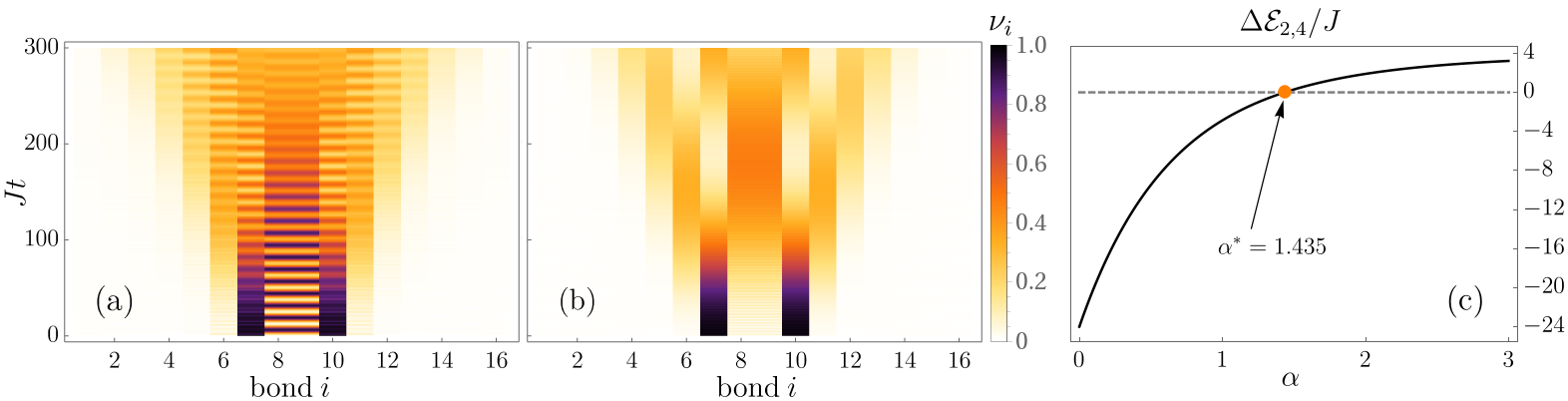}
 	\caption{Real-time dynamics of string breaking in the long-range Ising model. (a) String breaking as a two-stage process when the exponent $\alpha=1.435$ is  such that a resonance condition is satisfied. (b) String breaking also takes place with an non-resonant exponent $\alpha=1.1$. Note the lack of transient string oscillations, as opposed to the resonant case. Both instances show the domain-wall dynamics $\nu_i(t)$, for a system of size $L=17$, and $\ell=3$, $h_x/J=0.25$. (c) Graphical solution of the resonance condition for the example in (a). $\Delta \mathcal{E}_{2,4}$ is the energy difference (in units of $J$) between the initial state with two kinks and a four-kink state, in which the central spin is flipped.}
 	\label{fig:3}
 \end{figure*}

We start by outlining our main results, which will be analyzed in more detail in the following sections. We show the characteristic patterns of string breaking for the short-range and long-range quantum Ising chains in Figs.~\ref{fig:1}, \ref{fig:2} and Fig.~\ref{fig:3}, respectively. As a central observation, the phenomenon of string breaking takes place as a two-stage process. In the first stage, the two kinks remain essentially static, while the connecting string can become a dynamical object, see, in particular, Figs.~\ref{fig:2}(a) and \ref{fig:3}(a). We find that in the short-range Ising chain the stability of the initial kinks crucially depends on their initial distance $\ell$. Upon changing separation from $\ell=4$, Fig.~\ref{fig:1}, to $\ell=6$, Figs.~\ref{fig:2}(a) and \ref{fig:2}(b), the time range of their stability jumps from a time $Jt \approx 40$ to a value which is not anymore visible on the accessed time scales. It is one of the main goals of this work to provide a physical picture for this stability and to describe the string motion in this regime.

While the two initial kinks can remain stable for a long time, we observe that the connecting string can undergo complex dynamics, see Figs.~\ref{fig:2}(a) and \ref{fig:2}(b), in particular, which we explain in more detail via an effective description in Sec.~\ref{effective}. Especially for the case of long-lived initial kinks, particle-antiparticle pairs are created and annihilated in a complex oscillatory pattern without being able to induce a breaking of the string. Conversely, outside of the initial string, the system remains almost inert with only some slight dynamics induced by the quench such as the ballistic motion of a bound pair of two domain walls, an analog of a meson, in Fig.~\ref{fig:2}(a). Finally, we find that there are also parameter regimes where the string does not display dynamics during the initial stage, see Fig.~\ref{fig:2}(c) and the short-time behavior in Fig.~\ref{fig:3}(b). This latter feature will also be captured in our effective model, which shows that the dynamics of the string is too constrained in this case to induce oscillatory behavior. 
%Further, we also find parameter regimes for which the string not only remains immobile, but also does not break, see Figs.~\ref{fig:2}(c) and \ref{fig:2}(d).

A further important finding of this work is representatively shown in Fig.~\ref{fig:2}(b). During the first stage, the dynamics in the string not only generates particles but also significant quantum correlations, while these are absent outside of the central region, yielding a characteristic spatiotemporal correlation pattern. Consequently, the recently observed entanglement growth during string breaking in gauge theories~\cite{ PhysRevX.6.011023, QEDMPS, PhysRevX.10.021041, Magnifico2020realtimedynamics, PhysRevLett.124.180602} can be understood to be initially caused by the generation of these strong correlations inside of the string, while the outside remains effectively decoupled. Notice that this implies that the mesons traveling ballistically in Fig.~\ref{fig:2}(a), which are merely produced by the quench dynamics, are essentially decoupled from the inside of the string.

\section{First stage: string motion} \label{first}

Let us now focus on the first stage of string breaking. As a central observation, the two initial kinks can remain static for long times, which allows us to develop simplified effective descriptions in this regime.

Importantly, for the short-range Ising chain the system effectively decomposes into three disconnected spatial regions, in particular, because no quantum correlations are generated between them, see Fig.~\ref{fig:2}(b). Since the outside essentially remains static, we will now focus on the dynamics of the string itself, which in this decomposition, is now an object with a fixed spatial extent determined by the initial spin configuration and therefore the initial spatial separation $\ell$ of the kinks. Specifically, we will describe the string dynamics in the following by the Hamiltonians in Eqs.~(\ref{Ising_short},\ref{Ising_long}) on a chain of $\ell$ sites and initial condition $|\Psi_0\rangle= \bigotimes_{m=1}^\ell |\downarrow \rangle_m$. Let us point out that we have to impose a magnetic boundary condition at the ends, since the first and last spin of the string have to remain inert due to the requirement that the two initial domain walls are static. This can be achieved by skipping the transverse-field term or by adding a strong longitudinal field to those lattice sites.

For the long-range model, an analogous decomposition is not possible. However, we still observe that the spatial region outside of the initial string remains almost inert. Therefore, one can develop an effective description which keeps the spins outside of the string frozen and the spins inside the string as dynamical objects.

\subsection{Effective description of the string dynamics} \label{effective}

As argued in Sec.~\ref{Hamiltonians}, the considered quantum Ising chains exhibit confinement dynamics whenever $h_x \ll J$ and therefore whenever quantum fluctuations are weak. We take this as a starting point to organize the Hilbert space for the string dynamics. Specifically, we will decompose the state space into sectors with different numbers of domain walls. For that purpose, we introduce operators $\mathcal{P}_k$ projecting onto the subspace of $k$ kinks. This allows us to represent the effective Hamiltonian $\mathcal{H}_\mathrm{eff}$ for the string as:
\begin{equation}
\label{Heff}
\mathcal{H}_{\text{eff}}=\sum_{k \in \mathcal{I}} \mathcal{H}_{k} + \sum_{k \not= k'} \mathcal{V}_{k,k'}\, ,
\end{equation}
where $\mathcal{H}_k=\mathcal{P}_k H\mathcal{P}_k$ denotes the projection of the full Hamiltonians in Eqs.~\eqref{Ising_short} and \eqref{Ising_long} onto the subspace with $k$ kinks. Accordingly, $\mathcal{V}_{k,k'}=\mathcal{P}_kH\mathcal{P}_{k'} +\mathcal{P}_{k'}H\mathcal{P}_k$, stands for the coupling between such subspaces and $\mathcal{I}=\{2,4,\dots,k_\mathrm{max}\}$ is the index set labeling the allowed kink sectors up to the number of kinks $k_\mathrm{max}$, that maximally fit into the string upon respecting the boundary condition, which is $k_\mathrm{max}=\ell-2$ when $\ell$ is even and $k_\mathrm{max}=\ell-1$ when $\ell$ is odd. 

By decomposing the Hamiltonian into these kink sectors one obtains a representation as depicted in Fig.~\ref{fig:4}(a) for the short-range model in Eq.~(\ref{Ising_short}). The overall picture, however, does not change for the long-range case. The general structure of $\mathcal{H}_k$ can be divided into a diagonal part $\mathcal{E}_{k}$ in the spin configurations and an off-diagonal one, which is proportional to $h_x$ and acts as a hopping term for the kinks. The transitions between different kink sectors contained in $\mathcal{V}_{k,k'}$ are driven by single-spin flips induced by the transverse field, which can only connect spin configurations that differ by exactly two domain walls. In Appendix \ref{AppendixA}, we show explicitly how to construct all the different terms in Eq.~\eqref{Heff}.

Let us now more specifically analyze the structure of the diagonal part of $\mathcal{H}_k$. For the short-range model, it reads:
 \begin{equation}
 \mathcal{E}_{k}(\mathcal{S})=-J(\ell-1) + 2Jk -h_z(\ell-2l_{\mathcal{S}}),
 \label{energy_short}
 \end{equation}
 where $k$ and $l_{\mathcal{S}}$ denote the number of kinks and the number of $\downarrow$-spins in the given spin configuration $\mathcal{S}$, respectively. The sector of $k=0$ kinks only contains one configuration $\mathcal{S} =\mid\downarrow \dots \downarrow \rangle$, i.e., the initial condition. Since kinks can only be generated in pairs, the next higher sector is the $k=2$ one. The respective two domain walls can reside on various different bonds with an energy that depends linearly on their distance, which is the defining feature of confinement and which leads to a tower of states as depicted in Fig.~\ref{fig:4}(a), similarly also for the higher kink sectors. 
 
For the long-range system, the energy of a particular spin configuration is not a simple explicit function of the parameters $\ell$, $k$, and $l_{\mathcal{S}}$~\cite{Long-range}. Instead, we numerically obtain the energy for a given kink sector using the formula
\begin{equation}
    \mathcal{E}_k(\mathcal{S})= -J\sum_{i < j}^L \frac{s_i(\mathcal{S}) s_j(\mathcal{S})}{r_{ij}^{\alpha}},
    \label{long_energy}
\end{equation} 
where $s_i(\mathcal{S})=\pm 1$ is the value of the spin on site $i$ corresponding to the configuration $\mathcal{S}$. Unlike the short-range case, this will depend on $\ell$, $k$, and $l_{\mathcal{S}}$ nonlinearly, and notably the influence of boundary effects can be significant, as a general feature of long-range systems.

Transitions induced by the transverse field across configurations that live in a given sector (and therefore leave the number of domain walls invariant) have the only consequence that they move domain walls between %nearest-
neighboring lattice sites. As the domain walls are confined, such a motion always costs energy so that the respective process is off-resonant and therefore only yields perturbative corrections. For the short-range model, this can be alternatively seen by recognizing that the diagonal part $\mathcal{E}_k$ resembles a Wannier-Stark ladder of charged particles in an electric field~\cite{Wannier1960}  as a function of both $k$ and $l_{\mathcal{S}}$. Here, the role of the field is taken over either by the coupling $J$ or the longitudinal field $h_z$. The off-diagonal part of $\mathcal{H}_k$ induces motion on this Wannier-Stark ladder for a fixed $k$ via $h_x$ by flipping individual spins. As known from the Wannier-Stark problem, however, this motion is always off-resonant and therefore only slightly perturbs the eigenstates of $\mathcal{E}_{k}$. This holds, in particular, in the limit of weak kinetic energy, which is guaranteed in our problem as $h_x \ll J$, see the discussion in Sec.~\ref{Models}. It will therefore be sufficient for the moment to ignore this motion within sectors of a given number of kinks $k$.

Similar representations of Hamiltonians in kink sectors have been introduced and used for the effective description of systems with confinement~\cite{Coldea177, Long-range, Mazza, Rutkevich_2010}. Here, however, we not only restrict to low-kink sectors as in previous works but rather consider the full decomposition. As we will show, this turns out to be important for the description of the string dynamics because many resonant spin configurations $\mathcal{S}$ can appear across different kink sectors, which become crucial to  describe the string motion.

At this point, it becomes important to distinguish two different classes of parameter sets. Depending on the choice of Hamiltonian parameters, spin configurations in the higher-kink sectors can either be off-resonant or degenerate with the initial string. This distinction, which determines whether higher-kink sectors contribute perturbatively or nonperturbatively to the string dynamics,  will become crucial to identify situations where string motion is suppressed or induced, as explained below. For the short-range model,  resonances can  occur whenever
\begin{equation}
\label{commensurate}
   \frac{h_z}{J}=\frac{k}{\ell -l_\mathcal{S}},
\end{equation}

\noindent where $1+k/2 \le l_\mathcal{S} \le \ell-k/2$. In the case of the long-range interacting model, the resonance condition corresponds to matching the energy for two different configurations $\mathcal{E}_{k}(\mathcal{S}_1)= \mathcal{E}_{k'}(\mathcal{S}_2)$. 
The location of the resonance can be easily determined by numerically comparing the energy difference between kink sectors; see Fig.~\ref{fig:3}(c) for a particular example of tuning $\alpha$ to obtain a degeneracy. Notice that by taking into account the off-diagonal transverse-field contributions within fixed kink sectors, the energy levels in Eq.~(\ref{energy_short}) get broadened so that the resonance condition does not require fine-tuning.

When the parameters are such that there are no resonances, the string becomes inert and only acquires perturbative corrections from higher-kink sectors. An example of such a scenario is shown in Fig.~\ref{fig:3}(b) for the long-range model, where not only the initial charges remain static but also the string is almost inactive. Note, however, that in this example the string eventually breaks. The situation changes drastically in the short-range model where only a slight departure away from the resonance condition yields a suppression of both string motion and string breaking, at least, up to the accessible time scales, see Fig.~\ref{fig:2}(c) in comparison to Fig.~\ref{fig:1}(a). While the suppression of transport and particle production in the non-resonant short-range model were recently reported~\cite{Mazza,roeck2019slow}, here we also find that the spreading of quantum information is drastically reduced in the off-resonant case as compared to the resonant one, see Fig.~\ref{fig:2}(d) where the dynamics of the half-chain entanglement entropy is shown. 

Regarding the resonant case, which is illustrated in Figs.~\ref{fig:1}, \ref{fig:2}(a), \ref{fig:2}(b), and \ref{fig:3}(a), the situation is again completely different,  since the string can develop complex motion. Importantly, this dynamics is dominantly driven by all those spin configurations across all kink sectors which are resonant with the initial string configuration, as we will show in Sec.~\ref{resonant_subspace}.

It might appear as a fine-tuning problem to achieve resonant configurations. However, let us now argue that the resonant case is at least as generic as the off-resonant one. First of all, the absence of a resonance we attribute to a lattice effect. Due to a nonzero lattice spacing, the energy in the string develops a granular structure allowing only discrete values. This changes when going towards a continuum limit where this granularity is gradually washed out. Therefore, for small lattice spacings resonances become much more likely. Furthermore, notice that taking into account the broadening of the energy levels due to quantum fluctuations in $\mathcal{H}_k$ makes the resonance conditions more generic.

In the following, we will analyze the implications of the effective model in more detail by first deriving a bound on particle creation in the string and second by analyzing the dynamics in the resonant subspace.
 
\subsection{Bound on particle production} \label{part_prod_bound}

As emphasized before, the string dynamics is dominated by the resonant subspaces across the different kink sectors. This immediately has an important consequence: there always exists a maximum kink sector $k^\ast$ that is resonantly coupled to the initial string. This imposes a constraint on the number of domain walls $K$ that can be generated during real-time evolution. 

For the short-range model, we find that $K$ is bounded by 
 \begin{equation}
 K \le k^\ast = \bigg \lfloor \frac{2(\ell-1)}{1+2J/h_z} \bigg \rfloor_\mathrm{even}\, ,
 \label{bound}
 \end{equation}
\noindent where $2J/h_z$ is one of the rational numbers allowed by Eq.~\eqref{commensurate} and is also such that $k^\ast$ is at least equal to 2. Here, the notation $\lfloor x \rfloor_\mathrm{even}$ stands for the largest even integer smaller than or equal to $x$. Importantly, $k^\ast \le k_\mathrm{max}$ can be much smaller than the maximum number of kinks $k_\mathrm{max}$ that fit in a string of given length $\ell$ ignoring the resonance condition, especially upon decreasing the value of the longitudinal field $h_z$ where $k^\ast \propto h_z/J$ implying a small kink density. As anticipated before, $k_\mathrm{max}=\ell-2$ if $\ell$ is even and $k_\mathrm{max}=\ell-1$ when $\ell$ is odd.

\begin{figure}[tb!]
 	\centering
    \includegraphics[width=1.\columnwidth]{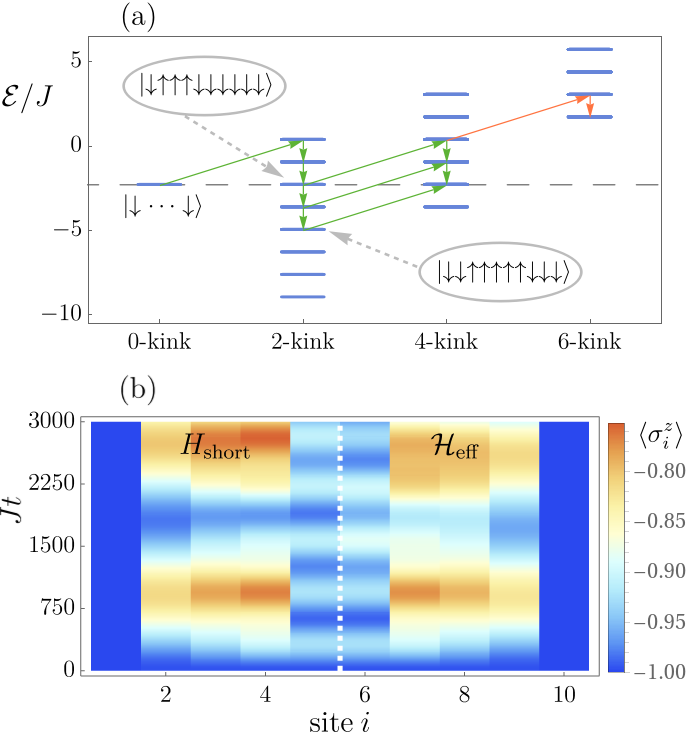}
 	\caption{Effective description of the string motion in the short-range model during the first stage. (a) Decomposition of the effective Hamiltonian $\mathcal{H}_{\text{eff}}$ into different kink sectors. The arrows indicate allowed transitions induced by  single-spin flips via the transverse field. Paths yielding virtual transitions between resonant states are shown in green. Orange arrows indicate transitions towards a non-resonant sector. Two spin configurations within the two-kink sector are shown, as an example, as well as their location in the energy ladder. (b) Comparison of the exact magnetization dynamics $\sigma_i^z(t)$  (left) to the effective description including all kink sectors up to $k^
 	\ast$ (right), with $\ell=10$, $h_x/J=0.075$, and $h_z/J=2/3$. }
 	\label{fig:4}
\end{figure}

We derive the bound of Eq.~(\ref{bound}) in Appendix \ref{AppendixB}. The origin of this bound can, however, be directly understood from Fig.~\ref{fig:4} where we depict the structure of the energy levels for the short-range model. The creation of two new kinks costs at least an energy $4J$. As a consequence, the minimum energy at a given kink sector has to increase for higher $k$ up to the point where the tower is shifted out of resonance, which marks the maximum number of domain walls which can be potentially generated. Of course, these considerations neglect the influence of off-diagonal spin flips in $\mathcal{H}_k$ so that the bound only holds in the limit of weak transverse fields and might yield corrections for larger $h_x.$ The derived bound represents a constraint on the generation of new kinks, which restricts the formation of composite mesonic objects of bound domain wall pairs and hence might significantly slow down string breaking.  

One particular implication of this bound is a controlled criterion for truncating the sums in Eq.~\eqref{Heff} incorporating all non-perturbative effects, i.e., what is the maximum kink sector that has to be taken into account for the description of the string dynamics. In order to assess and illustrate the approach presented here, in Fig.~\ref{fig:4}(b) we show the dynamics of the mean on-site magnetization $\langle \sigma_i^z(t)\rangle$, of a string of length $\ell=10$ in a longitudinal field $h_z/J=2/3$, a value for which the resonance condition is met. As implied by the bound~\eqref{bound} and shown in Fig.~\ref{fig:4}(a), here $k^\ast=4$, so that the corresponding reduced model simply reads $\mathcal{H}_{\text{eff}}=\mathcal{H}_0 +\mathcal{H}_2+\mathcal{H}_4 +\mathcal{V}_{0,2} +\mathcal{V}_{2,4}$. As can be observed, the reduced model captures the main features of the exact dynamics.

 For the long-range model, it is, in principle, possible to get more than two states with resonant energies after we impose  $\mathcal{E}_{k}(\mathcal{S}_1)= \mathcal{E}_{k'}(\mathcal{S}_2)$ and choose $\alpha$ accordingly. However, due to the nonlinear nature of the energy function of  the long-range model, given by Eq.\ (\ref{long_energy}), it becomes more challenging to get a strict bound on the number of resonantly accessible domain walls. Let us point out that, it is, however, still possible to determine \emph{numerically} the maximum kink sector simply by scanning the energy in Eq.~\eqref{long_energy} in all relevant kink sectors to identify degeneracies with the initial string state.

\subsection{Dynamics in resonant subspace} \label{resonant_subspace}
 
The full solution of the Hamiltonian in Eq.~(\ref{Heff}) may still require exponential resources in the string length $\ell$. Here, we aim to show that a further reduction is possible beyond the restriction onto the maximum kink sector $k^\ast$ that has been taken into account already in the previous Section. Specifically, it is possible to obtain an effective description of the resonant subspace alone, which as we show provides further insights during the first stage of the string dynamics.

The central property that we will use in the following is that all spin configurations outside of the resonant subspace can be treated perturbatively in $h_x/J$, by recalling that the transitions between spin configurations are driven by the transverse field, which has to be chosen to satisfy $h_x \ll J$. %for our quantum spin chains. 
However, in general, the challenge is that, in principle, exponentially many paths exist in the energy level diagram such as in Fig.~\ref{fig:4}(a) that can connect different resonant configurations by virtual transitions. It is clear, nonetheless, that those paths that require overcoming large energy differences are less relevant than the others. It turns out that the identification of the ``shortest paths'' that are contributing dominantly depends on the details of the chosen parameters.

For the short-range model, we indicate in Fig.~\ref{fig:4}(a) with arrows the shortest paths in the energy diagram for one particular case of a string of length $\ell=10$ with $h_z/J=2/3$ connecting dominantly the different resonant sectors in terms of single-spin flips. We can then ignore all states not contained in this shortest paths selection, since they will only contribute subdominantly yielding only further perturbative corrections. The remaining off-resonant spin configurations can then be eliminated perturbatively by means of a Schrieffer-Wolff transformation \cite{SW, Muhlschlegel1968}, as explained in Appendix \ref{AppendixC}. This yields an effective theory for the resonant subspace alone. We applied this approach to the string of the example shown in Fig.~\ref{fig:1}. One can show (see Appendix \ref{AppendixC}) that the effective model for the resonant states, in this case, maps onto a two-level system. Hence, one can predict analytically the time scale at which the new kinks are generated in the string. This happens when the spin configuration with two kinks in the interior of the string is maximally populated for the first time. According to our model, this occurs at $Jt^* = \pi/(2(h_x/J)^2) \approx 39.3$, which is in excellent agreement with the results shown in Fig.~\ref{fig:1}. In Appendix~\ref{AppendixTimescales}, we investigate the accuracy of this prediction at increasing transverse-field strength. Let us already note at this point that the present analysis also has central implications for the second stage of string breaking, that will be discussed in the following Section.  

In the long-range model, the resonant dynamics are especially simple because the resonant subspace only contains  two states. For the particular case chosen in Fig.~\ref{fig:3}(a), the transition between the two states requires  flipping only one spin in the center of the string. In such a case, the oscillation period between these two states can be directly calculated.  As shown in Appendix \ref{AppendixC}, the time at which the higher-kink state inside the string is maximally populated is $Jt^* = \pi/(h_x/J) \approx 12.57$ for the parameters used in Fig.~\ref{fig:3}(a), which is in perfect agreement with ED results of the many-body Hamiltonian.

Let us emphasize that the analytical estimates of the typical time scales for the onset of string breaking, which are obtained with our effective description, go beyond the estimates for the non-resonant scenario as reported in Ref.~\cite{roeck2019slow}.

\section{Second stage: string breaking} \label{secondStage}

While the final string breaking can be prolonged to long times, see Fig.~\ref{fig:2}, it is known especially for the short-range model that the system is ergodic and thermalizing~\cite{KimHuse}, although long-lived nonequilibrium states have been recently discussed in this system~\cite{2011Banuls, 2019James, 2019Robinson} and delayed thermalization observed in  the long-range model~\cite{Neyenhuise1700672}. However, in general, we expect that the considered models will eventually restore a homogeneous state where the string has to be broken. For the case displayed in Fig.~\ref{fig:1}, we indeed observe that at long times the system becomes homogeneous with some remaining spatiotemporal fluctuations expected for systems of finite size~\cite{D'Alessio2016}.

Eventually, the string breaks by the formation of mesons, i.e., bound pairs of domain walls involving, in particular, the two initial kinks. Strings can, in principle, break both for the case of resonant motion, see Figs.~\ref{fig:1}(a) and~\ref{fig:3}(a), as well as when the parameters are chosen such that the resonance condition for the string motion is not satisfied, see Fig.~\ref{fig:3}(b). The latter case seems to be especially applicable  to the long-range model, since when the resonance condition is not met in the short-range model, string breaking may only occur after an exponentially long time, see Fig.~\ref{fig:2}(c) and Refs.~\cite{Mazza, roeck2019slow}.  
Furthermore, we also observe another significant difference between the long- and short-range models. While the time scale of string breaking does not seem to depend crucially on varying the parameters for the long-range interacting case, see Fig.~\ref{fig:3}, for the short-range Ising chain string breaking can be delayed by orders of magnitude in Fig.~\ref{fig:2}(a) by only changing the initial string length from $\ell=4$ to $\ell=6$.

As we aim to argue in the following, the delayed string breaking and meson formation for large string lengths $\ell$ in the short-range model is not only caused by the energy costs for particle creation due to the large kink mass as in the Schwinger mechanism~\cite{Sauter1931, Schwinger1951}. We rather observe that there are, in particular, strong kinetic constraints imposed by the dynamics in the resonant subspace. First of all, the considerations from Sec.~\ref{resonant_subspace} imply that only a limited subset of spin and therefore domain wall configurations is kinetically accessible. In this context, we find that there are mainly two different scenarios.

On the one hand, the resonant subspace might be such that a configuration with newly generated domain walls close to the initial kinks can be reached. This makes the meson formation very efficient. Such a case is displayed in Fig.~\ref{fig:1}, where we find that the time scale for string breaking coincides with the time scale of reaching the respective resonant domain wall configuration. In Sec.~\ref{resonant_subspace},  we have discussed that from the effective description the latter time scale is $Jt^* = \pi/(2(h_x/J)^2) \approx 39.3$ matching the data in Fig.~\ref{fig:1} obtained using exact diagonalization.

On the other hand, the resonant subspace can induce kinetic constraints so that only domain walls at larger distances from the initial kinks can be generated. In this context, the general bound on domain wall production derived in Eq.~(\ref{bound}) provides some general implications. In particular, for weak symmetry-breaking fields $h_z$ the maximally accessible kink density in the string becomes proportional to $h_z/J$ implying that the typical distance between the generated domain walls is large. This makes it difficult for the system to efficiently form mesons of two kinks at a short separation. 
%long-range 

For the long-range case, only a single higher-kink configuration can be resonant with the initial string,  unless we fine-tune multiple parameters. As a consequence, we have not identified a case where the time scales associated with kink dynamics and string breaking have been related to each other. For the resonant case displayed in Fig.~\ref{fig:3}(a), this explains why there are a large number of oscillations before string breaking, which is analogous to what is seen for the short-range case displayed in Fig.~\ref{fig:2}(a). In addition, for generic parameters, the long-range model has no resonances and string breaking occurs with no transient string oscillations, as shown in Fig.~\ref{fig:3}(b). For the short-range model, the minimal energy gap between two spin configurations is always a constant value, see Fig.~\ref{fig:4}(a). However, for the long-range Hamiltonian, due to the nonlinear nature of the energy expression Eq.~\eqref{long_energy}, the  spacing between higher energy states can be extremely small. Due to this nature, the string can still break relatively fast, even without satisfying a resonant condition, see Fig.~\ref{fig:3}(b).

\section{Concluding discussion} \label{Conclusions}

In this work, we have shown that string breaking can occur dynamically in quantum Ising chains where domain walls develop a confining potential induced either by a symmetry-breaking longitudinal field~\cite{mccoy, Rutkevich2008, calabreseNat} or by long-range interactions~\cite{Long-range, SilvaLongRange}. Our main observation is that this phenomenon can be described as a two-stage process. During the first stage, a pair of initial kinks effectively acts as static external charges. The connecting string, however, can become a dynamical object and develops complex dynamics. To approximate this dynamics, we have derived %for whose real-time evolution we derive 
an effective kinetically constrained model in the resonant subspace. In particular, we have obtained a bound on the maximal number of kinks that can be dynamically generated, and, for some cases, obtained a quantitative estimate for the time scale of final string breaking. We have argued that the large time scales for eventual string breaking are not only caused by the energy costs for pair creation due to the large mass of particles as in the Schwinger mechanism~\cite{Sauter1931, Schwinger1951}. We rather find that the effective model in the resonant subspace also imposes strong kinetic constraints. In this context, a natural question is to what extent the observed slow string breaking dynamics can be related to the slow relaxation observed previously in kinetically constrained models~\cite{PhysRevB.92.100305, Prem2017,  PhysRevLett.121.040603, PhysRevLett.120.030601, bulmash2018generalized, PhysRevResearch.2.012003, PhysRevX.10.011047,  Pai2019, PhysRevB.101.174204, PhysRevB.100.214313, PhysRevB.101.125126, Bernien2017, Turner2018, Turner2018PRB, PhysRevX.10.021041}. In this respect, the non-resonant local dynamics in the short-range model seems to be even more constrained, with both particle production and spreading of quantum information being strongly suppressed.

While all of our analysis has been carried out for quantum Ising models, it can be equally well applied also to lattice gauge theories. For instance, it might be particularly interesting to explore the constrained dynamics in the resonant subspaces for such systems, as well as the string stability after a quench, as a function of the separation. A further interesting route might be the extension of our analysis to string breaking dynamics in higher-dimensional systems, which is certainly much more challenging. Importantly, long strings or flux tubes connecting far distant static background charges can still behave as effectively one-dimensional~\cite{1981Luescher}, which might make our analysis also applicable in this case and therefore relevant for high-energy physics.

Our finding, that the phenomenology of string breaking dynamics cannot only be realized in gauge theories but also in systems with less complexity such as spin chains, implies that this phenomenon might be more directly accessible experimentally. The dynamics in spin chains has already been successfully studied in various quantum simulator experiments~\cite{2016Smith,2016Martinez,2017Zeiher,Bernien2017,2017Monroe53,2017Jurcevic,2018Browaeys,2018Bakr,deLeseleuc2018,Marcuzzi2017}, while lattice gauge theories are much more challenging to realize, as gauge invariance is difficult to enforce, with, however, some notable recent efforts~\cite{2016Martinez,natPhys2018lgt1,Mil1128,natPhys2019lgt3}. More specifically, we now outline how our results might be observable in Rydberg atom and trapped ion quantum simulators within the current scope of technology. Both platforms support, in principle, the initial preparation of any targeted product state~\cite{Marcuzzi2017, 2017Jurcevic, 2017Monroe53} such as those with two domain walls, as depicted in Fig.~\ref{fig:1}(b). Since the strength of next-nearest-neighbor interactions in Rydberg atoms is just about $1.6\%$ of the nearest-neighbor value~\cite{2017Zeiher}, it is safe to neglect interactions beyond nearest neighbors up to timescales $Jt \sim 100$. Therefore, this type of platform can be used to probe short-range Ising chains~\cite{2017Zeiher,Bernien2017,2018Browaeys,2018Bakr,deLeseleuc2018,Marcuzzi2017} as in Eq.~(\ref{Ising_short}), up to the mentioned timescales. On the other hand, long-range interacting Ising models find a natural implementation in systems of trapped ions with a tunable power-law exponent~\cite{2016Smith,2017Jurcevic,2017Monroe53}. However, the timescales necessary for the observation of string breaking in the numerical data we show in this work are rather large compared to what has been achieved experimentally. Importantly, these timescales can be significantly tuned by increasing the transverse-field strength $h_x$, as long as $h_x$ does not exceed a critical value beyond which domain walls cease to be elementary excitations of the Ising model; see Appendix~\ref{AppendixTimescales}. We emphasize that, even in the regime of strong transverse-field, where domain walls can no longer be regarded as point-like particles, one can use a field-theoretical approach to take into account the finite width of kinks. As shown in Appendix~\ref{AppendixE}, this yields a similar description to the one at weak fields. On the other hand, what might be certainly experimentally observable is the constrained dynamics during the first stage where interesting and complex dynamical patterns are realized, see Fig.~\ref{fig:2}. Moreover, both considered experimental platforms allow for local readouts which make all the quantities discussed in this work measurable.  

Finally, let us remark that, although our effective models allow us to elucidate various interesting aspects of  the first stage of string breaking, and even to predict  typical timescales for the final breaking of the string, a \emph{complete} understanding of the second stage remains a challenge for techniques relying on classical resources. In this sense, the experimental perspectives with quantum simulators discussed above are crucial, as it is this approach that stands as the most promising route for deepening our understanding of hard problems such as string breaking dynamics, in a foreseeable future.

\textit{Note added} Recently, we became aware of a related complementary work on confinement-induced quasilocalized dynamics \cite{lerose2019quasilocalized}. Also, two experimental works appeared~\cite{2019arXiv191211117T, 2020arXiv200103044V}, constituting the first experimental realization of the real-time dynamics of confinement in Ising chains with quantum simulators. These works clearly demonstrate the feasibility to implement experimentally both the models and the initial condition  herein considered. Yet, the observation of string breaking dynamics in Ising chains remains an interesting goal for future experiments.

\begin{acknowledgments}

We acknowledge discussions with  G.S.\ Bali, D.\ Banerjee, D.\ Barredo, P.\ Becker, J.\ Berges, P.\ Bienias, A.\ Browaeys, K.\ Collins, M.\ Dalmonte,  A.\ De, G.\ Dvali, L.\ Feng, A.\ Gambassi,  H.B.\ Kaplan, A.\ Kyprianidis,  A. Lerose,  R.\ Lundgren,  C.\ Monroe, G.\ Pagano, J.\ Preskill, and  W.L.\ Tan.
F.L., S.W., and A.V.G. acknowledge funding by the NSF PFCQC program, DoE BES Materials and Chemical Sciences Research
for Quantum Information Science program (award No.\ DE-SC0019449), DoE ASCR Accelerated Research in Quantum
Computing program (award No.\  DE-SC0020312), DoE ASCR Quantum Testbed Pathfinder program (award No.\ DE-SC0019040), AFOSR, AFOSR MURI, ARO MURI, ARL CDQI, and NSF PFC at JQI. S.W. acknowledges support from the NIST NRC Postdoctoral Associateship award. This project has received funding from the European Research Council (ERC) under the European Union’s Horizon 2020 research and innovation programme 
(grant agreement No. 853443), and M.H. further acknowledges support by the 
Deutsche Forschungsgemeinschaft via the Gottfried Wilhelm Leibniz Prize 
program. R.V. thanks the Galileo Galilei Institute for Theoretical Physics for the hospitality during the workshop ``\emph{Breakdown of Ergodicity in Isolated Quantum
Systems: From Glassiness to Localization}'' and the INFN for partial support during the completion of this work.

\end{acknowledgments}

\appendix

\section{Construction of the effective Hamiltonian $H_\text{eff}$} \label{AppendixA}

In this Appendix, we explicitly show how to construct the different terms that appear in Eq.~\eqref{Heff} of the main text. Here we regard the short-range model, although a   derivation for the long-range one can, in principle, be done analogously.  Let us recall that here we consider a chain of length $\ell$, with  a magnetic boundary condition imposed at the ends, and that the reference state is the initial string, that is,  $|\Psi_0\rangle= \bigotimes_{m=1}^\ell |\!\downarrow \rangle_m$.

The projected model in the 0-kink sector reads
\begin{equation}
\label{H0}
\mathcal{H}_0=\mathcal{P}_0H\mathcal{P}_0=\mathcal{E}_0 |\Psi_0\rangle\langle\Psi_0|,
\end{equation}
\noindent where $\mathcal{P}_0=|\Psi_0\rangle\langle\Psi_0|$ is the projector onto the 0-kink sector, and $\mathcal{E}_0 = -J(\ell-1)+\ell h_z$ is the energy of the initial string.

Let us now look at the two- and four-kink sectors. Elements of the two-kink subspace are labeled by two quantum numbers $j_1$ and $j_2$, such that $|j_1,j_2\rangle=|\!\downarrow\cdots \downarrow_{j_1}\uparrow\cdots \uparrow \downarrow_{j_2} \cdots \downarrow\rangle$, with $j_1=1, \dots, \ell-2$ and $j_2=j_1+2, \dots, \ell$. The two-kink projected Hamiltonian acts  on  $|j_1,j_2\rangle$ as
\begin{align}
\label{H2}
\mathcal{H}_2|j_1,j_2\rangle&=\mathcal{E}_2(j_1,j_2) |j_1,j_2\rangle - h_x[|j_1+1,j_2\rangle \nonumber \\ & +|j_1-1,j_2\rangle
+|j_1,j_2+1\rangle+|j_1,j_2-1\rangle],
\end{align}
\noindent where the diagonal term is $\mathcal{E}_2=-J(\ell-5)-h_z[\ell-2(l_1+l_2)]$, and $l_1=j_1$, $l_2=\ell-j_2+1$ are the lengths of the two resulting strings. 

The four-kink model requires four quantum numbers: $|j_1,j_2, l_2, j_3\rangle=|\!\downarrow\cdots \downarrow_{j_1}\uparrow\cdots \uparrow \downarrow_{j_2} \cdots \downarrow_{(j_2+l_2-1)}\uparrow \cdots \uparrow \downarrow_{j_3} \cdots \downarrow\rangle$, with indices taking the possible values $j_1=1, \dots,\ell-4$, $j_2=j_1+2, \dots, \ell-2$, $j_3=j_2+2, \dots, \ell$, $l_2=1, \dots, j_3-j_2-1$. The action of the four-kink projected Hamiltonian on $|j_1,j_2, l_2, j_3\rangle$ is given by
\begin{align}
\label{H4}
\mathcal{H}_4|&j_1,j_2, l_2, j_3\rangle=\mathcal{E}_4(j_1,j_2, l_2, j_3) |j_1,j_2, l_2, j_3\rangle \nonumber \\ &- h_x\big[|j_1+1,j_2, l_2, j_3\rangle+|j_1-1,j_2, l_2, j_3\rangle \nonumber \\ & |j_1,j_2+1, l_2-1, j_3\rangle +|j_1,j_2-1, l_2+1, j_3\rangle \nonumber \\ &|j_1,j_2, l_2+1, j_3\rangle+|j_1,j_2, l_2-1, j_3\rangle \nonumber \\ & |j_1,j_2, l_2, j_3+1\rangle+|j_1,j_2, l_2, j_3-1\rangle\big],
\end{align}
\noindent where $\mathcal{E}_4(j_1,j_2, l_2, j_3) =-J(\ell-9)-h_z[\ell-2(l_1+l_2+l_3)]$, and $l_1=j_1$, $l_3=\ell-j_3+1$.

The off-diagonal elements of these projected Hamiltonians act as effective  hopping terms for the kinks. Yet, in order to fully account for string breaking, we need to take into account the couplings between sectors. Such transitions are induced by the transverse field and are given by 
\begin{equation}
\label{V02}
\mathcal{V}_{0 \to 2}=\mathcal{P}_2H\mathcal{P}_0=-h_x\sum_{j_1=1}^{\ell-2}|j_1,j_1+2\rangle \langle\Psi_0 |,
\end{equation}
\noindent and likewise for $\mathcal{V}_{2 \to 0}$, so that $\mathcal{V}_{0, 2}=\mathcal{V}_{0 \to 2}+\mathcal{V}_{2 \to 0}$.  Analogously, the coupling between the two- and four-kink subspaces is given by
\begin{align}
\label{V24}
\mathcal{V}_{2 \to 4}&=\mathcal{P}_4H\mathcal{P}_2 \nonumber \\  &=-h_x\Bigg[\sum_{j_1=1}^{\ell-4}\sum_{j_2=j_1+2}^{\ell-2}\sum_{j_3=j_2+2}^{\ell}|j_1,j_2,1, j_3\rangle \langle j_1,j_3 |\nonumber \\  &+\sum_{j_1=1}^{\ell-4}\sum_{j_3=j_1+4}^{\ell}\sum_{l_2=1}^{j_3-j_1-3}|j_1,j_1+2,l_2, j_3\rangle \langle j_1+l_2+1,j_3 |\nonumber \\ &+\sum_{j_1=1}^{\ell-4}\sum_{j_2=j_1+2}^{\ell-2}\sum_{j_3=j_2+2}^{\ell}|j_1,j_2,j_3-j_2-1, j_3\rangle \langle j_1,j_2 | \Bigg],
\end{align}
\noindent and likewise for $\mathcal{V}_{4 \to 2}$.  Terms involving a higher number of kinks can be derived in a similar manner.

\section{Derivation of Eq.~\eqref{bound}} \label{AppendixB}

Here we derive a bound on the maximum number of kinks that can be resonantly produced in a string of length $\ell$, when considering the short-range model. In this case the leading-order energy of a spin configuration is given by Eq.~\eqref{energy_short}, whereas the energy of the initial string is $\mathcal{E}_0 =-J(\ell-1)+\ell h_z$. Imposing the resonance condition $\mathcal{E}_0\equiv \mathcal{E}(\mathcal{S})$ and solving for $k$, yields
\begin{equation}
\label{appen1}
k=\frac{h_z}{J}(\ell-l_{\mathcal{S}}).
\end{equation}

On the other hand,  $k$ and $l_{\mathcal{S}}$ are not independent. Indeed, one can readily show that 
\begin{equation}
\label{appen2}
l_{\mathcal{S}}\ge l_{\mathcal{S}}^{\text{min}}(k)=1+\frac{k}{2}.
\end{equation}

Combining \eqref{appen1} and \eqref{appen2} gives a bound,  in terms of $\ell$ and $h_z$,  on the number of kinks $K$ that is possible to produce inside the string, provided that the resonance condition is met, namely
 \begin{equation}
 K \le k^\ast = \bigg \lfloor \frac{2(\ell-1)}{1+2J/h_z} \bigg \rfloor_\mathrm{even}\,,
 \end{equation}
\noindent  where $\lfloor x \rfloor_\mathrm{even}$ gives the  largest even integer smaller than or equal to $x$. Note that the value of $h_z/J$ must also be consistent with the resonance condition.

\section{Effective model for the resonant subspace via a Schrieffer-Wolff transformation } \label{AppendixC}

Here we show how to construct an effective description for the resonant states of the reduced model via a Schrieffer-Wolff (SW) transformation \cite{SW, Muhlschlegel1968}. As explained in the main text, one starts by identifying the shortest paths connecting the resonant states in adjacent kink sectors, as illustrated in Fig.~\ref{fig:4}(a). Then we neglect those states that do not form part of such paths. Next we apply the SW transformation to eliminate the remaining off-resonant states and generate effective couplings between resonant configurations in neighboring sectors. 

This approach works as follows. Let us consider a Hamiltonian of the form
\begin{equation}
\label{C1}
H= H_0 +\lambda V,
\end{equation}
\noindent where the eigenvalues $\{\epsilon_{\mu}\}$ and eigenstates $\{|\mu\rangle\}$ of $H_0$ are known,  that is,
\begin{equation}
\label{C2}
H_0 = \sum_{\mu} \epsilon_{\mu} |\mu\rangle\langle\mu|.
\end{equation}

Our task is to carry out a unitary transformation, with generator $S$, such that the rotated Hamiltonian 
\begin{align}
\label{C3}
\tilde{H} &= \text{e}^{S}(H_0+\lambda V)\text{e}^{-S} \nonumber \\
&=H_0 +\lambda V +[S, H_0] +\lambda[S,V] +\frac{1}{2}[S,[S,H_0]]+\cdots
\end{align}
\noindent has no off-diagonal terms to first order. This is accomplished by choosing $S$ such that 
\begin{equation}
\label{C4}
[S,H_0] =-\lambda V,
\end{equation}
\noindent that is, 
\begin{equation}
\label{C7}
S_{\mu\nu}=\lambda \frac{V_{\mu\nu}}{\epsilon_{\mu}-\epsilon_{\nu}},
\end{equation}
\noindent in the $H_0$ eigenbasis, and provided that the right-hand side is finite. Thus, Eq.~\eqref{C3} becomes
\begin{equation}
\label{C5}
\tilde{H} = H_0 +\frac{1}{2}\lambda[S,V]+\mathcal{O}(\lambda^3).
\end{equation}

Now let us apply this technique to the problem at hand. The starting point is the effective Hamiltonian $\mathcal{H}_{\text{eff}}$ as given in Eq.~\eqref{Heff}, including contributions from the $k$ sector with $k$ as big as needed. The projected Hamiltonians $\mathcal{H}_{k}$ contain a diagonal part, given essentially by Eq.~\eqref{energy_short}, and an off-diagonal part. Here we collect  the diagonal terms in $H_0$ and put all of the off-diagonal contributions (all having to do with the transverse field $h_x$) in the perturbation $\lambda V$, with $h_x/J$ playing the role of the small parameter $\lambda$. Note that, after performing the SW transformation, the second-order term, $\frac{1}{2}\lambda[S,V]$, will contain  effective couplings between resonant states in adjacent kink subspaces, which are then used to build the effective theory. 
\begin{figure}[tb!]
 	\centering
 		\includegraphics[width=\columnwidth]{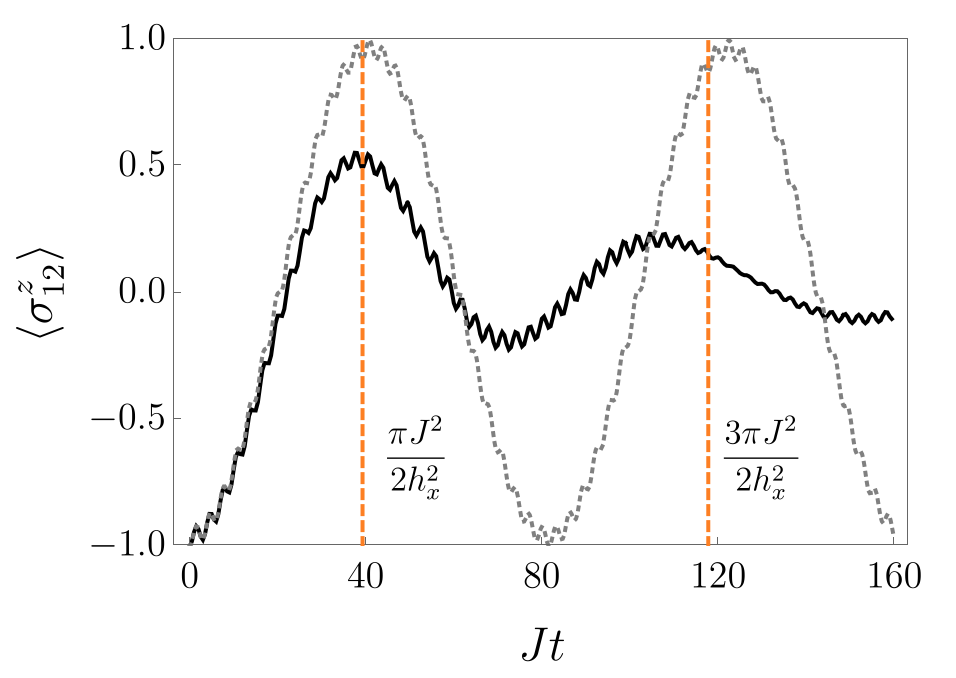}
 	\caption{ Dynamics of mean local magnetization at site 12 $\langle \sigma_{12}^z(t)\rangle$, in a chain of size $L=24$, with a central string of length $\ell=4$,  for the short-range model~\eqref{Ising_short} with $h_x/J=0.2, h_z/J=1$. The black curve shows the complete solution obtained by ED, whereas the gray dotted curve is the perturbative solution where we only consider the string alone with fixed kinks at the boundaries (also computed via ED). The difference between the orange dashed lines gives the period as predicted by the effective model for the resonant states.}
 	\label{fig:AppC1}
\end{figure}

 \begin{figure}[t]
 	\centering
 		\includegraphics[width=\columnwidth]{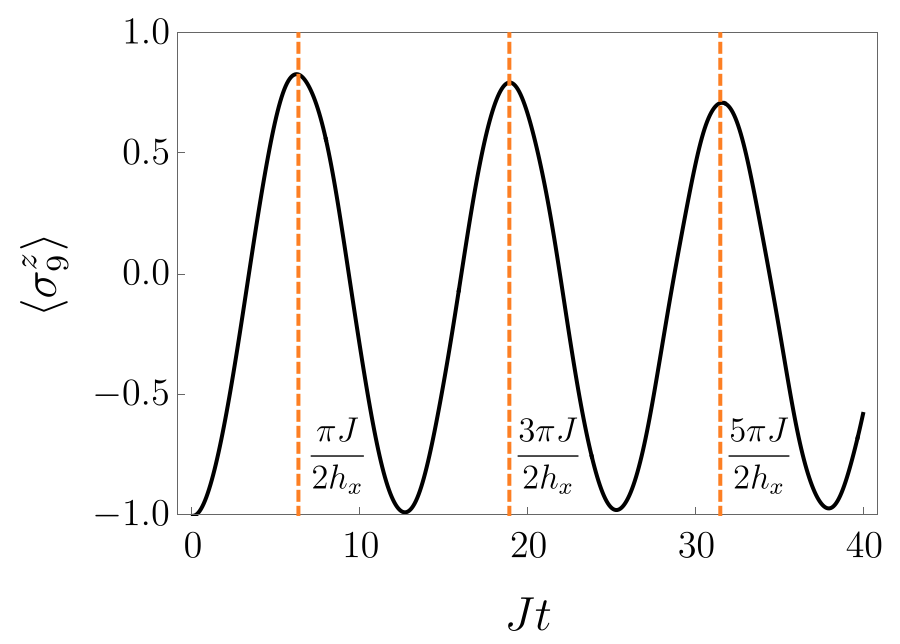}
 	\caption{Dynamics of mean local magnetization at site 9 $\langle \sigma_9^z(t)\rangle$, within a string of length $\ell=3$, embedded in the center of a chain with $L=17$ spins, for the long-range model~\eqref{Ising_long} with $h_x/J=0.25, \alpha=1.435$. The black curve is the complete solution obtained by ED. The difference between two consecutive orange dashed lines gives the period as predicted by the effective model for the resonant states.}
 	\label{fig:AppC2}
\end{figure} 

   \begin{figure*}[t!]
 	\centering
 		\includegraphics[width=\textwidth]{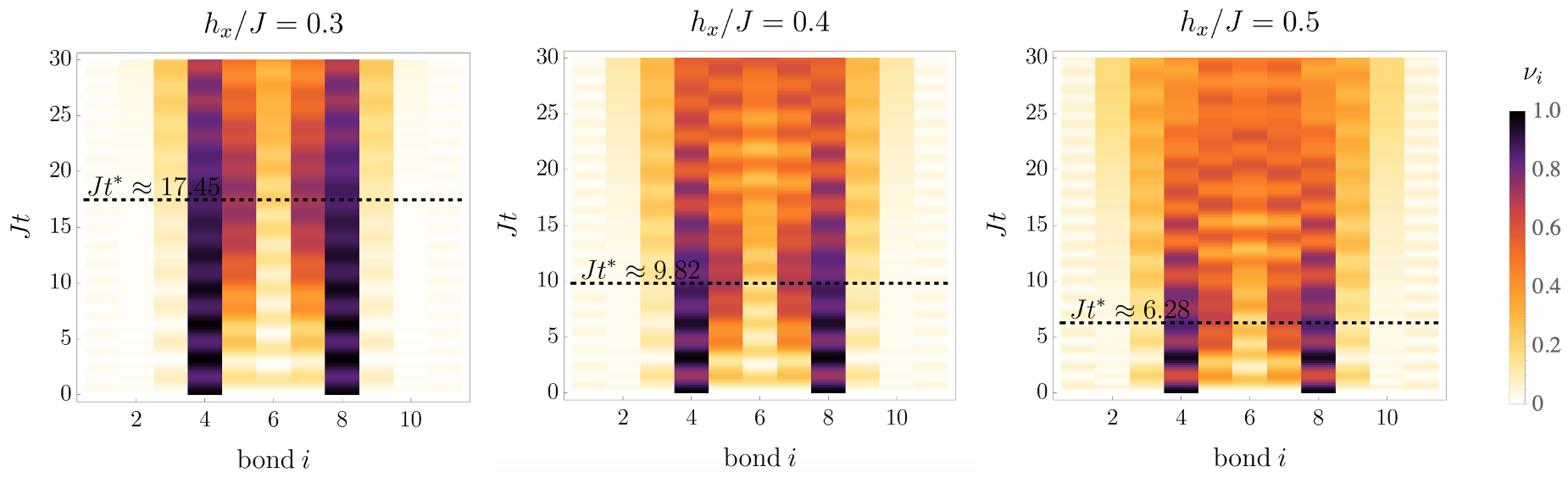}
 	\caption{Timescales for the onset of string breaking in the short-range Ising model with $L=12, h_z/J=1$ and an initial distance $\ell=4$ between two domain walls, at increasing transverse-field strength $h_x/J=0.3, 0.4$, and $0.5$. The black dashed lines indicate the predicted typical timescale from the effective model in the resonant subspace.}
 	\label{fig:AppTimescales}
 \end{figure*}
Applying this method to the string of the example in Fig.~\ref{fig:1}, we get that the effective Hamiltonian for the two resonant states $\mid\Psi_0\rangle=\mid\downarrow\downarrow\downarrow\downarrow\rangle$ and $\mid\Psi_1\rangle=\mid\downarrow\uparrow\uparrow\downarrow\rangle$, is akin to the one of a two-level system, that is,
\begin{equation}
\label{C8}
h_{\text{eff}}=\begin{pmatrix} 
a & b \\
b & c 
\end{pmatrix}
\end{equation}
\noindent with $a/J = -2 \frac{(h_x/J)^2}{4-2h_z/J}$, $b/J = -\frac{(h_x/J)^2}{2h_z/J}-\frac{(h_x/J)^2}{4-2h_z/J} $, and $c/J = -2 \frac{(h_x/J)^2}{2h_z/J}$. Then it becomes a simple matter to determine the period of the oscillations between the two resonant states, namely, $JT=\pi/(h_x/J)^2$ (see Fig.~\ref{fig:AppC1}).

A effective theory for the resonant sector can also be constructed for long-range interacting  systems, as illustrated in Fig.~\ref{fig:AppC2}, where we show the example of a string of length $\ell=3$ embedded in a overall chain of size $L=17$, corresponding to the case discussed in Fig.~\ref{fig:3}(a). Here, our effective description predicts a period of $JT=\pi/(h_x/J))\approx 12.57$, which is in perfect agreement with the exact results.  

\section{Timescales for the observation of string breaking}\label{AppendixTimescales}

As stated in the concluding remarks, the timescales necessary for the observation of string breaking can be tuned upon increasing the strength of the transverse field. In this Appendix, we illustrate this point concretely for the short-range model. Moreover, we employ the effective model in the resonant subspace discussed above to obtain an analytical prediction for the relevant timescale.   Thus, let us consider a similar setting as in Fig.~\ref{fig:1}, namely, we study the quench dynamics  to $h_x/J=0.2, h_z/J=1$,  in the short-range model, starting from a string state with a central string of length $\ell=4$, embedded in a chain of $L=12$ spins with open boundary conditions. As shown in the  Appendix~\ref{AppendixC}, for this particular setting, our effective description maps the problem onto a two-level system. The breaking of the string is thus expected when the ``broken-string'' configuration, $\mid\downarrow\uparrow\uparrow\downarrow\rangle$, is maximally populated for the first time, for that represents the moment when two new kinks are created right next to the original ones. According to our effective model, this occurs at a time $Jt^{*}=\frac{\pi}{ 2(h_x/J)^2}$. 

In Fig.~\ref{fig:AppTimescales}, we show the domain-wall dynamics obtained via ED at three different values of the transverse-field strength, namely, $h_x/J=0.3, 0.4$, and $0.5$, with the predicted typical timescales being $Jt^*\approx 17.45,  9.82$, and $6.28$, respectively.  We observe that these predicted values give a remarkably good estimate of the onset of string breaking, even at relatively large transverse fields. Overall, these results show that by increasing the strength of the transverse field, it could be, in principle, possible to bring the whole phenomenology of string breaking within the reach of current technologies in quantum simulators.

 \section{Extended kinks in the short-range Ising chain} \label{AppendixE}

 In the majority of this paper, we have worked close to the limit $h_x \approx 0$ for the short-range model, Eq.~\eqref{Ising_short}. Directly at $h_x = 0$, the kinks are given by product states, $\sigma^z_i = \pm 1$, but at $h_x >0$ single local kinks are generally complicated states which are not easily constructed, even with the exact analytic solutions to this Hamiltonian in the deconfined ($h_z = 0$) limit.

 In this Appendix, we construct approximate kinks which are still product states, but better approximate the actual kinks of the model.
A simple ansatz is to take
\beq
| \mathcal{K} \rangle = \bigotimes_{j} \Big[ \cos \theta_j | + \rangle + \sin \theta_j | - \rangle \Big],
\eeq
where $\sigma^z | \pm \rangle = \pm | \pm \rangle$, and we choose the expectation value on each site to be the position-space profile of the kinks:
\beq
\langle \mathcal{K} | \sigma^z_j | \mathcal{K} \rangle = \cos 2 \theta_j = F(x_j).
\eeq

Here, we take $F(x)$ to vanish at the position of the kink (which should be at the half-way point between two lattice sites). 
This equation only determines $\sin \theta_j$ up to a sign; if we are working with $h_x,J>0$, we should choose the sign which makes $\langle \sigma^x_j \rangle = \sin 2 \theta_j$ positive.
We furthermore want the state to approach the exact magnetization of the Ising chain far away from the position of any kinks:
\beq
\lim_{x \rightarrow \pm \infty} F(x) = \pm N_0.
\eeq
We can determine $N_0$ directly from the exact solution \cite{pfeuty70}, namely
\beq
N_0 = \left[ 1 - (h_x/J)^2 \right]^{1/8}.
\eeq
In addition, we would like the profile $F(x)$ to have a finite width around the position of the kink.
This width should be on the order of
\beq
W \sim \frac{v}{E_\mathcal{K}},
\eeq
where $v$ is a characteristic velocity in the system and $E_\mathcal{K}$ is the energy of the kink.
Here, we can take some intuition from the exact dispersion of the model,
\beq
\epsilon_\kappa = 2 \sqrt{J^2 + h_x^2 - 2 h_x J \cos \kappa},
\eeq
where $\kappa \in [-\pi,\pi]$.

 \begin{figure}[tb!]
 	\centering
 		\includegraphics[width=\columnwidth]{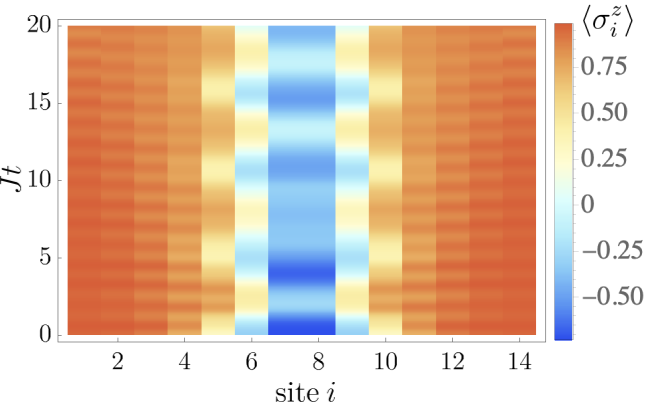}
 	\caption{Evolution of the mean on-site magnetization $\langle \sigma_i^z(t)\rangle$ in the short-range model~\eqref{Ising_short} with $h_x/J=0.45, h_z/J=0.6$, taking into account the approximate expression for the extended kinks [Eq.~\eqref{eq:ansatz}].}
 	\label{fig:AppD}
\end{figure}

So one sees that the energy gap of the system is $\Delta = 2 (J - h_x)$, and that the low-momentum dispersion takes the form $\sqrt{\Delta + v^2 \kappa^2} \approx \Delta + v^2 \kappa^2/2\Delta$, identifying $v = 2 \sqrt{h_x J}$. So in principle we could build any function $F(x)$ which approaches $\pm N_0$ away from positions of kinks and vanishes along a width of order $W \sim \sqrt{h_x/J}/(1 - h_x/J)$ in units of the lattice spacing.

We propose the following single-kink ansatz:
\beq
F(x) = \pm N_0 \tanh \left[ \frac{3 E_\mathcal{K}}{4 N_0^2 v} (x - x_0) \right].
\label{eq:ansatz}
\eeq
This functional form was inspired by the semiclassical static solution for kinks in scalar $\phi^4$ quantum field theory (QFT). This QFT describes the short-range Ising chain at couplings $h_x/J$ such that the correlation length is large compared with the lattice spacing, and furthermore we can treat this theory perturbatively in its interactions if we are not too close to the phase transition ($h_x = J$).
In these limits, the kinks are given by the field configurations of Eq.~\eqref{eq:ansatz} \cite{Rajaraman1982}.
This expression also reduces to a step function in the limit $h_x \rightarrow 0$, as expected. The expression $E_\mathcal{K}$ is the exact energy of the kink, but at our level of approximation we simply take $E_\mathcal{K} = \Delta$. 
Numerical results using these approximate kinks are shown in Fig.~\ref{fig:AppD}, where we take $h_x/J=0.45$ and $h_z/J=0.6$. We find similar behavior to that seen in the small $h_x$ limit, where the string undergoes dynamical oscillations, but the location of the initial kinks remains approximately static. In principle, one may perform similar experiments for higher transverse fields in the long-range model after numerically obtaining reasonable values for $N_0$, $v$, and $E_\mathcal{K}$ and a similar ansatz to the above, although we have not done so here. This provides some evidence that the dynamics described in the main text of this paper survives into the many-body regime, where $h_x$ is not small.

\bibliography{BIB_1}

\end{document}